\documentclass[aip,jcp,reprint]{revtex4-1}
\usepackage{epsfig}
\usepackage{graphicx}
\usepackage{amsmath}
\usepackage{amssymb}
\usepackage{amsfonts}  
\usepackage{fancyhdr}
\usepackage{bm}
\usepackage{dcolumn}
\usepackage{subfigure}
\newcommand{\intmat}{\mathbf{\mathcal{M}}}
\newcommand{\bra}[1]{\langle #1|}
\newcommand{\ket}[1]{|#1\rangle}
\newcommand{\braket}[2]{\langle #1|#2\rangle}
\newcommand{\dsum}{\displaystyle\sum}
\newcommand{\half}{\frac{1}{2}}
\newcommand{\rvec}{\mathbf{R}}
\newcommand{\rvecp}{\mathbf{R}^\prime}

\newcommand{\rveco}{\mathbf{R}_0}
\newcommand{\rvecop}{\mathbf{R}^\prime_0}
\newcommand{\Pvec}{\mathbf{P}}
\newcommand{\Pveco}{\mathbf{P}_0}
\newcommand{\Pvecp}{\mathbf{P}^\prime}

\newcommand{\Pvecop}{\mathbf{P}^\prime_0}
\newcommand{\Pvecopt}{\mathbf{P}^{\prime\;T}_0}
\newcommand{\xvec}{\mathbf{x}}
\newcommand{\xveco}{\mathbf{x}_0}
\newcommand{\xvecp}{\mathbf{x}^\prime}

\newcommand{\xvecpt}{\mathbf{x}^{\prime\;T}}
\newcommand{\xvecop}{\mathbf{x}^\prime_0}
\newcommand{\xvecopt}{\mathbf{x}^{\prime\;T}_0}
\newcommand{\pvec}{\mathbf{p}}
\newcommand{\pveco}{\mathbf{p}_0}
\newcommand{\pvecp}{\mathbf{p}^\prime}

\newcommand{\pvecpt}{\mathbf{p}^{\prime\;T}}
\newcommand{\pvecop}{\mathbf{p}^\prime_0}
\newcommand{\pvecopt}{\mathbf{p}^{\prime\;T}_0}

\newcommand{\zvec}{\mathbf{z}}
\newcommand{\proj}{\mathcal{P}}
\newcommand{\hop}{H}

\newcommand{\bytwo}[1]{\frac{#1}{2}}

\begin{document}
\title{Exact quantum statistics for electronically 
nonadiabatic systems using continuous path variables}
\author{Nandini Ananth}
\author{Thomas F. Miller III}
\affiliation{Division of Chemistry and Chemical Engineering, 
California Institute of Technology, Pasadena, California 91125, USA}
\date{\today}
\begin{abstract}
   We derive an exact, continuous-variable path integral 
   (PI) representation of the canonical partition function 
   for electronically nonadiabatic systems.
   Utilizing the Stock-Thoss (ST) mapping for an $N$-level system, 
   matrix elements of the Boltzmann operator are expressed in 
   Cartesian coordinates for both the nuclear and 
   electronic degrees of freedom.
   The PI discretization presented here properly constrains 
   the electronic Cartesian coordinates
   to the physical subspace of the mapping.
   %[G. Stock and M. Thoss,Phys. Rev. Lett. 78, 578 (1997)].
   We numerically demonstrate that the resulting PI-ST representation 
   is exact for the calculation of equilibrium properties of 
   systems with coupled electronic and nuclear degrees of freedom.
   We further show that the PI-ST formulation 
   provides a natural means to initialize semiclassical 
   trajectories for the calculation of real-time 
   thermal correlation functions, which is numerically 
   demonstrated in applications to a series of nonadiabatic model systems.
\end{abstract}
\maketitle
\section{Introduction}
Electronically nonadiabatic processes lie at the heart
of  chemical phenomena, including electron solvation dynamics,
\cite{erb95,fw91,bs91,bs92,am09,tfm08}
energy transfer at metal surfaces,\cite{amw04} and radiationless
transitions in the condensed phase.\cite{hn08,fkf76}
Elucidating the mechanisms and rates for these processes
remains a critical challenge from both the theoretical 
and experimental perspectives.

In an electronically nonadiabatic process, the Born-Oppenheimer 
separation of nuclear and electronic motions breaks down, 
%necessitating a description of the feedback between them.
necessitating a description of the coupled motions of the electrons and nuclei.
The exponential scaling of exact quantum mechanical methods has
motivated the development of numerous mixed quantum-classical (MQC) 
methods for nonadiabatic dynamics, in which the nuclei are typically 
treated using classical mechanics and electronic degrees of freedom (DoF) 
are treated at the quantum mechanical level.
These methods, which include the broad classes of mean-field\cite{pe27,cz04} 
and surface hopping\cite{pp69,rk99,ad00,jct90,ovp97,awj02,mfh95,yhw07}
approaches, have been succesfully employed in a range of applications.
However, it has been shown that processes including non-radiative electronic 
relaxation\cite{er98} and resonance energy transfer\cite{whm09} require
a consistent description of the coupling between the electronic and nuclear DoF,
an inherently challenging task in the MQC framework.\cite{jcb00}

Semiclassical (SC) methods allow for a dynamically consistent
treatment of the electronic and nuclear motion.\cite{whm09,hdm79,gs97,gs05}
This can be achieved by mapping discrete electronic states 
to a continuous variable representation using, for instance, 
spin coherent states\cite{yt75,jrk79,hk80,ms00}
or bosonization
techniques based on angular momentum theory.\cite{th40,js65}
In particular, Stock and Thoss used the latter approach to 
derive the mapping Hamiltonian,
%the Meyer-Miller-Stock-Thoss (MMST) or mapping Hamiltonian,
an exact Cartesian representation of the quantum Hamiltonian 
for an $N$-level system.\cite{gs97} %% READDRESS
The mapping Hamiltonian has been successfully used in the SC
description of processes for which a system initially occupies 
a pure electronic state.\cite{xs97,sb051,sb052,na07} 
However, use of this approach to calculate real-time 
quantum thermal correlation functions (TCFs)  has relied
on initializing SC trajectories to an approximate description
of the Boltzmann distribution.\cite{hw98,jap03,eb09}
The demonstrated sensitivity \cite{aag00,gs99,gs05} of the 
calculated TCFs to the initialization scheme 
indicates that progress is needed to improve the 
accuracy and generality of this approach.

In this paper, we derive an exact continuous variable PI 
representation for the Boltzmann distribution of general,
N-level systems.
This is achieved in the mapping framework using a projection 
operator to constrain the electronic coordinates to the physical 
subspace of the mapping.
We numerically demonstrate that the resulting PI-ST representation is 
exact for the calculation of equilibrium properties of two- and
three-state systems with coupled electronic and nuclear DoF.
We further show that the PI-ST formulation can be used to 
initialize SC trajectories to an exact quantum Boltzmann 
distribution, and using the SC initial value representation 
(SC-IVR) method,\cite{whm01} we obtain accurate real-time TCFs for a 
series of nonadiabatic model systems.
%%fakesection : pimmst derivation
\section{Theory}
\subsection{The mapping Hamiltonian}
Consider the general $N$-level Hamiltonian operator
\begin{equation}
	H = h_0(\rvec,\Pvec) + 
	\dsum_{n,m=1}^N V_{nm}(\rvec)\ket{\psi_n}\bra{\psi_m},
	\label{eq:diab_ham}
\end{equation}
where $(\rvec,\Pvec)$ represent the nuclear positions and momenta, 
$\{\psi_n\}$ is the basis of electronic states, and
%(the adiabatic representation can also be used) 
$\{V_{nm}(\rvec)\}$ is the set of potential energy
matrix elements.
Furthermore, $h_0(\rvec,\Pvec)=T(\Pvec) + V_0(\rvec)$,
where $T(\Pvec)$ is the nuclear kinetic energy operator
and $V_0(\rvec)$ is a state-independent part of the potential
energy.

Following the ST mapping approach,\cite{gs97}
the $N$-level system is represented by a system of $N$ uncoupled harmonic 
oscillators (HO), such that 
\begin{eqnarray}
	\ket{\psi_n}\bra{\psi_m}&\rightarrow& a_n^+a_m\\
	\ket{\psi_n}&\rightarrow&\ket{0_10_2\cdots 1_n\cdots 0_N}.
	\label{eq:st_mapping}
\end{eqnarray}
Here, we have introduced the boson creation and annihilation
operators $a_n^+$ and $a_n$, which obey the commutation rules
$\left[ a_n^+,a_m \right]=\delta_{nm}$. We have also introduced the
 singly excited
oscillator (SEO) states $\ket{n}\equiv\ket{0_1\cdots 1_n\cdots 0_N}$,
which are $N$-oscillator eigenstates with a single quantum of excitation in the
n$^{th}$ mode. % has a single quantum of excitation.
%(include commutation relations?)
The resulting form of the Hamiltonian operator is
\begin{equation}
   H = h_0(\rvec,\Pvec)+\dsum_{n,m=1}^N a_n^+ V_{nm}(\rvec) a_m,
\label{eq:bosonham}
\end{equation}
or equivalently in the SEO basis,
\begin{equation}
	H=h_0(\rvec,\Pvec) + \sum_{n,m=1}^N \ket{n}V_{nm}(\rvec)\bra{m}.
	\label{eq:seo_ham}
\end{equation}
Introducing the Cartesian representation of the 
boson operators,
\begin{equation}
	{x}_n = \frac{1}{\sqrt{2}}(a_n + a_n^+)
	\;\;\text{and}\;\;
	{p}_n = \frac{i}{\sqrt{2}}(a_n^+ - a_n),
	\label{eq:intro_xp}
\end{equation}
we obtain the corresponding Cartesian representation
of the Hamiltonian operator,
\begin{equation}
	H = h_0(\rvec,\Pvec) + 
	\frac{1}{2}\dsum_{n,m=1}^N\left({x}_n{x}_m+
	{p}_n{p}_m -\delta_{nm}\right)V_{nm}(\rvec).
	\label{eq:mmst_ham}
\end{equation}
The mapping Hamiltonian in Eq.~(\ref{eq:mmst_ham}), also known as the Meyer-Miller-Stock-Thoss Hamiltonian,
was originally derived as a classical model for an electronically nonadiabatic system;\cite{hdm79} 
it was later shown to be an exact and general representation for the quantum mechanical Hamiltonian.\cite{gs97}
%mechanical mapping by Stock and Thoss.\cite{gs97}

\subsection{PI discretization}
The canonical partition function is obtained from 
the trace of the Boltzmann operator,
\begin{equation}
	Z = \text{Tr}\left[ e^{-\beta \hop} \right],
\end{equation}
where $\beta$ is the reciprocal temperature and the 
trace is taken over the states that span the
electronic and nuclear DoF.
The resolution of the identity for this space can be 
expressed as
\begin{equation}
	\mathbf{I} = \int d\rvec \dsum_{n=1}^N 
	   \ket{\rvec,n}\bra{\rvec,n},
	\label{eq:seo_idres}
\end{equation}
where $\rvec$ indicates nuclear positions
and $n$ indicates the SEO state, as before.
Repeated insertion of this completeness relation
yields a path integral discretization of the 
partition function,
\begin{eqnarray}
       Z&=&\\
       \nonumber
       &&
       \int d\{\rvec_\alpha\} 
	\sum_{\{n_\alpha\}=1}^N
	\prod_{\alpha=1}^P
	\bra{\rvec_\alpha n_\alpha} e^{-\beta_P\hop}
	\ket{\rvec_{\alpha+1} n_{\alpha+1}},\;\;\;
	\label{eq:mmst_pirep}
\end{eqnarray}
where $P$ is the number of time-slices and $\beta_P=\beta/P$.
We have introduced the notation $\int d\left\{\rvec_\alpha\right\}
\equiv\left( \prod_{\alpha=1}^P\int d\rvec_\alpha\right)$ and 
$\sum_{\{n_\alpha\}=1}^N\equiv\left( \prod_{\alpha=1}^P\sum_{n_\alpha=1}^N\right)$.

The standard Trotter approximation\cite{et58}
can be used to factorize the matrix elements 
in this equation, yielding
\begin{eqnarray}
	\nonumber
	Z&=&\lim_{P\rightarrow\infty}\int d\left\{\rvec_\alpha
	\right\}
	\prod_{\alpha=1}^P 
	\left(\frac{MP}{2\pi\beta} \right)^\frac{f}{2}
	e^{-\beta_P V_0(\rvec_\alpha)}\\
	\nonumber
	&\times&
	\text{exp}\left[-\frac{MP}{2\beta}
	(\rvec_\alpha-\rvec_{\alpha+1})^T
	\cdot(\rvec_\alpha-\rvec_{\alpha+1})\right] \\
	&\times&\sum_{\{n_\alpha\}=1}^N\;
	\prod_{\alpha=1}^P
	\bra{n_\alpha}e^{-\beta_P\mathcal{V}(\rvec_\alpha)}
	\ket{n_{\alpha+1}},
	 \label{eq:nuc_trot}
\end{eqnarray}
where $f$ is the number of nuclear DoF,
$\mathcal{V}(\rvec)=\sum_{n,m=1}^N\ket{n}V_{nm}(\rvec)\bra{m}$
is the potential energy operator, and we set $\hbar=1$
throughout this paper.
In Eq.~(\ref{eq:nuc_trot}), we have assumed that the 
Hamiltonian in Eq.~(\ref{eq:diab_ham}) is expressed in the 
diabatic representation, although 
a similar approach could also be pursued in the adiabatic 
representation.\cite{hdm79}
Similar expressions for PI discretization in the 
basis of discrete electronic states have been obtained.
\cite{cds99,mha01,jrs07}

The SEO basis in Eq.~(\ref{eq:nuc_trot}) can be transformed to the Cartesian 
coordinate basis by first using a projection operator to select 
the subset of SEO states from the full set of N-oscillator states,
\begin{equation}
	\sum_{n=1}^N \ket{n}\bra{n}  =  \prod_{i=1}^N\left[ 
	\sum_{j_i=0}^\infty \ket{j_i}\bra{j_i}\right]\proj,
	\label{eq:seo_osc}
\end{equation}
where the projectior operator is defined as
\begin{equation}
	\proj=\dsum_{n=1}^N\ket{n}\bra{n}.
	\label{eq:proj_def}
\end{equation}
Then, the full oscillator basis is replaced using
\begin{equation}
	\sum_{j_i=0}^\infty \ket{j_i}\bra{j_i}
	=\int d\text{x}_i\ket{\text{x}_i}\bra{\text{x}_i},
	\label{eq:osc_coord}
\end{equation}
yielding a transformation from the SEO basis to
the Cartesian coordinate basis for the electronic DoF,
\begin{equation}
	\sum_{n=1}^N\ket{n}\bra{n}=
	   \int d\xvec \ket{\xvec}\bra{\xvec}\proj.
	\label{eq:seo_xproj}
\end{equation}
As in Klauder's work with spin coherent states,\cite{jrk97}
the projection operator in Eq.~(\ref{eq:seo_xproj}) constrains 
electronic coordinates to a specific manifold in phase space. 
However, PI formulations using spin coherent states have only proven 
numerically tractable in the semiclassical limit,
\cite{jrk85,jrk97,al99,ms00,aa00,vk07} 
whereas we shall derive a PI formulation that can be used for exact 
numerical simulations.

Substituting Eq.~(\ref{eq:seo_xproj}) into Eq.~(\ref{eq:nuc_trot}),
we obtain an exact PI discretization of the canonical partition
function in continuous variables,
\begin{eqnarray}
	\nonumber
	Z&=&\lim_{P\rightarrow\infty}
	\int d\left\{\rvec_\alpha\right\}
	\prod_{\alpha=1}^P 
	\left(\frac{MP}{2\pi\beta} \right)^\bytwo{f}
	e^{-\beta_P V_0(\rvec_\alpha)}\\
	\nonumber
	&\times&
	e^{-\frac{MP}{2\beta}(\rvec_\alpha-\rvec_{\alpha+1})^T.
	(\rvec_\alpha-\rvec_{\alpha+1})}
	\int d\left\{\xvec_\alpha\right\}
	\\
	&\times&
	\prod_{\alpha=1}^P
	\bra{\xvec_\alpha}
	e^{-\beta_P\mathcal{V}(\rvec_\alpha)}\proj
	\ket{\xvec_{\alpha+1}},
	\label{eq:cv_part}
\end{eqnarray}
leaving only the task of evaluating the matrix elements
in the last term to obtain a computationally useful expression.

We note that a short-time approximation to the electronic matrix
elements could be performed directly in the Cartesian
representation at this stage. However, we find that a more 
numerically stable result is obtained by making the approximation 
in the SEO representation, such that
\begin{eqnarray}
	\bra{\xvec}e^{-\beta_P \mathcal{V}(\rvec)}
	\proj\ket{\xvecp}
	=\displaystyle\sum_{n,m=1}^N\braket{\xvec}{n}
	\intmat_{nm}(\rvec)
	\braket{m}{\xvecp},
	\label{eq:step2_derivation}
\end{eqnarray}
where $\intmat_{nm}(\rvec)=\bra{n}e^{-\beta_P\mathcal{V}(\rvec)}\ket{m}$.
Recognizing that the coordinate space SEO wavefunction 
is the product of $(N-1)$ ground state HO
wavefunctions and one first excited state HO
wavefunction, we have
\begin{equation}
	\braket{\xvec}{n}=\frac{\sqrt{2}}{\pi^{N/4}}
	\;[\xvec]_n\;e^{-\half\xvec^T\cdot\xvec},
	\label{eq:seo_wvfn}
\end{equation}
where $\left[ . \right]_n$ denotes the $n^\text{th}$
component of the enclosed vector.
The Boltzmann matrix element in the SEO representation can then be obtained 
following textbook procedures,\cite{dc87} such that to order 
$\vartheta(\beta_P^2)$,
\begin{equation}
	\intmat_{nm}(\rvec)=\left\{
	   \begin{array}{cc}
	   e^{-\beta_P V_{nn}(\rvec)}&,n=m, \\
	   -\beta_P V_{nm}(\rvec)\;e^{-\beta_P V_{nn}(\rvec)}
	   &,n\ne m.\\
	   \end{array}\right.
	\label{eq:int_mat}
\end{equation}
In the zero coupling limit, the matrix elements 
$\intmat_{nm}(\rvec)$ assume a diagonal form so that the different 
components of the electronic position vectors do not mix.
The off-diagonal matrix elements are related to 
the penalty of ring-polymer kink formation, in which neighboring PI
time-slices reside on different diabatic electronic 
surfaces.\cite{dc81,aoc83}

Finally, substituting Eqs.~(\ref{eq:seo_wvfn}) and (\ref{eq:int_mat}) into
Eq.~(\ref{eq:cv_part}), we arrive at the exact, continuous-variable
PI-ST representation of the canonical partition function for a 
nonadiabatic system with $f$ nuclear 
DoF coupled to $N$ electronic states,
\begin{eqnarray}
	\nonumber
	Z &=& \lim_{P\to\infty}
	\left(\frac{2MP}{\beta\pi^{N+1}}\right)^\frac{fP}{2}
	\int d\left\{\rvec_\alpha \right\} 
	\int d\left\{\xvec_\alpha\right\}\\
	&\times&
	\prod_{\alpha=1}^P 
	\mathcal{A}_\alpha\;\mathcal{F}_\alpha\;\mathcal{G}_\alpha,
	\label{eq:pist}
\end{eqnarray}
where
\begin{eqnarray}
	&&\mathcal{A}_\alpha=
	e^{-\frac{MP}{2\beta}
	(\rvec_\alpha-\rvec_{\alpha+1})^T\cdot
	(\rvec_\alpha-\rvec_{\alpha+1})}\;
	e^{-\beta_PV_0(\rvec_\alpha)},
	\label{eq:formula_a}
	\\
	&&\mathcal{F}_\alpha
	=\xvec_\alpha^T\;
	\intmat(\rvec_\alpha)
	\;\xvec_{\alpha+1},\;\; \text{and}
	\label{eq:formula_f}
	\\
	&&\mathcal{G}_\alpha=
	e^{-\xvec_\alpha^T\cdot\xvec_\alpha}.
	\label{eq:formula_g}
\end{eqnarray}
Eqs. (\ref{eq:pist}) - (\ref{eq:formula_g})
will be used to calculate numerically exact equilibrium
properties for nonadiabatic systems.

We note that our PI-ST formulation is different from
the result derived in Ref.~\onlinecite{sb01}, since
we include a projection operator to constrain the
system to the physical subspace of the mapping.
For cases where the system is prepared in a pure
SEO state, this constraint is implicitly obeyed, and 
the two PI representations are equivalent.
However, treatment of Boltzmann distributed systems
require that the projection operator
be explicitly included, as is done in the present study.
 %%fakesection : equilibrium implementation:
\section{equilibrium simulations}
\subsection{Implementation details}
The equilibrium properties considered in the
current paper include the nuclear probability distribution, 
the state-specific nuclear probability distribution, and the 
average total energy.
All equilibrium simulations were performed using standard 
path integral Monte Carlo (PIMC) importance sampling techniques,
although the use of path integral molecular dynamics (PIMD)
methods is also straightforward with the formulation developed here.

\textit{The nuclear probability distribution}
is defined as
\begin{equation}
	P(\rvec)=
	\frac{\text{Tr}[\delta(\rvec-\hat{\rvec})
	e^{-\beta\hop}]}
	{\text{Tr}[e^{-\beta\hop}]}.
\end{equation}
Using the PI-ST representation for the Boltzmann
operator, we obtain
\begin{eqnarray}
	\nonumber
	P(\rvec)&&=
	\\
	&&
	\frac{\int d\{\rvec_\alpha\} 
	\int d\{\xvec_\alpha\} 
	\delta(\rvec-\rvec_P)\displaystyle\prod_{\alpha=1}^{P}
	\mathcal{A}_\alpha \mathcal{G}_\alpha \mathcal{F}_\alpha}
	{\int d\left\{\rvec_\alpha\right\} 
	\int \left\{d\xvec_\alpha\right\} 
	\displaystyle\prod_{\alpha=1}^P
	\mathcal{A}_\alpha \mathcal{G}_\alpha \mathcal{F}_\alpha},
	\;\;\;
	\label{eq:nuc_prob_distrib}
\end{eqnarray}
where $\mathcal{A}_\alpha$, $\mathcal{F}_\alpha$ and 
$\mathcal{G}_\alpha$
are defined in Eqs.~(\ref{eq:formula_a})~-~(\ref{eq:formula_g}).
Importance sampling can then be performed using
\begin{eqnarray}
	\nonumber
	W(\{\xvec_\alpha\},\{\rvec_\alpha\})&=&\displaystyle\prod_{\alpha=1}^P
	\mathcal{A}_\alpha
	\mathcal{G}_\alpha
	|\mathcal{F}_\alpha|,
	\label{eq:pist_sampling}
\end{eqnarray}
where the absolute value in the last term ensures
a non-negative sampling function.
The expression for the nuclear probability distribution
is thus
\begin{equation}
	P(\rvec) =\frac{\left<\;\delta(\rvec-\rvec_P)\text{sgn}(\mathcal{F})\;\right>_W}
	{\left<\;\text{sgn}(\mathcal{F})\;\right>_W},
	\label{eq:nuc_prob_est}
\end{equation}
where 
\begin{equation}
	\text{sgn}(\mathcal{F}) = \prod_{\alpha=1}^P 
	\mathcal{F}_\alpha/|\;\mathcal{F}_\alpha\;|,
	\label{eq:signf}
\end{equation}
and the angle brackets in Eq.~(\ref{eq:nuc_prob_est}) indicate
the ensemble average with respect to the distribution
$W(\{\xvec_\alpha\},\{\rvec_\alpha\})$.

\textit{The state-specific nuclear probability distribution}
is obtained 
from the projection of the nuclear probability distribution
onto a given electronic state,
\begin{eqnarray}
	\nonumber
	P(n,\rvec)&=& \frac{\text{Tr}[\delta(\rvec-\hat{\rvec})
	\ket{n}\bra{n} e^{-\beta\hop}]}
	{\text{Tr}[e^{-\beta\hop}]} \\
	&=& 
	\frac{\langle\; \delta(\rvec-\rvec_P)\tilde{\mathcal{F}_n}\;
	\text{sgn}(\mathcal{F})\;\rangle_W}
	{\langle\;\text{sgn}(\mathcal{F})\;\rangle_W},
	\label{eq:eqbm_popn}
\end{eqnarray}
where
\begin{equation}
	\tilde{\mathcal{F}_n}=
	\frac{
	\left[\xvec_P^T\intmat(\rvec_P)\xvec_1 \right]_n}
	{\xvec_P^T\intmat(\rvec_P)\xvec_1}.
\end{equation}
%The notation $\left[ . \right]_n$ is used to denote the $n^\text{th}$
%component of the enclosed vector and 
The elements of the matrix $\intmat(\rvec)$ are defined in Eq.~(\ref{eq:int_mat}).

\textit{The average total energy}
of the system is obtained using 
a primitive energy estimator,
\begin{eqnarray}
	\nonumber
	\langle E\rangle &=&
	-\frac{1}{Z}\frac{\partial Z}{\partial \beta}\\
	&=& 
	\frac{\langle\;(\frac{P}{2\beta} + \tilde{\mathcal{F}} -
	\frac{\partial \mathcal{A}}{\partial\beta})
	\;\text{sgn}(\mathcal{F})\;\rangle_W}
	{\langle\;\text{sgn}(\mathcal{F})\;\rangle_W},
	\label{eq:energy_est}
\end{eqnarray}
where
\begin{eqnarray}
	\tilde{\mathcal{F}}
	&=&\displaystyle\sum_{\alpha=1}^P \frac{\xvec_\alpha^T 
	\;\frac{-\partial \intmat(\rvec_\alpha)}{\partial \beta}\;\xvec_{\alpha+1}}
	{\xvec_\alpha^T\intmat(\rvec_\alpha)\xvec_{\alpha+1}},
\end{eqnarray}
and
\begin{eqnarray}
  \nonumber
   \frac{\partial \mathcal{A}} {\partial \beta} 
   &=& \displaystyle\sum_{\alpha=1}^P \left[\frac{MP}{2\beta^2}
   (\mathbf{R}_\alpha-\mathbf{R}_{\alpha+1})^T.
   (\mathbf{R}_\alpha-\mathbf{R}_{\alpha+1}) \right.\\
   &&\left. \hspace{1.5in} -
   \frac{1}{P}V_0(\mathbf{R}_\alpha)\right].
\end{eqnarray}
\subsection{Equilibrium simulation results}
%%fakesection : Models
The first model that we consider (model I) includes a
two-state system coupled to a single vibrational DoF; 
it is a standard benchmark for the treatment 
of equilibrium statistics in nonadiabatic systems.
\cite{jrs07,mha01}

% two state : potential
\begin{center}
\begin{figure}[!htb]
	\includegraphics[angle=0,scale=0.35]{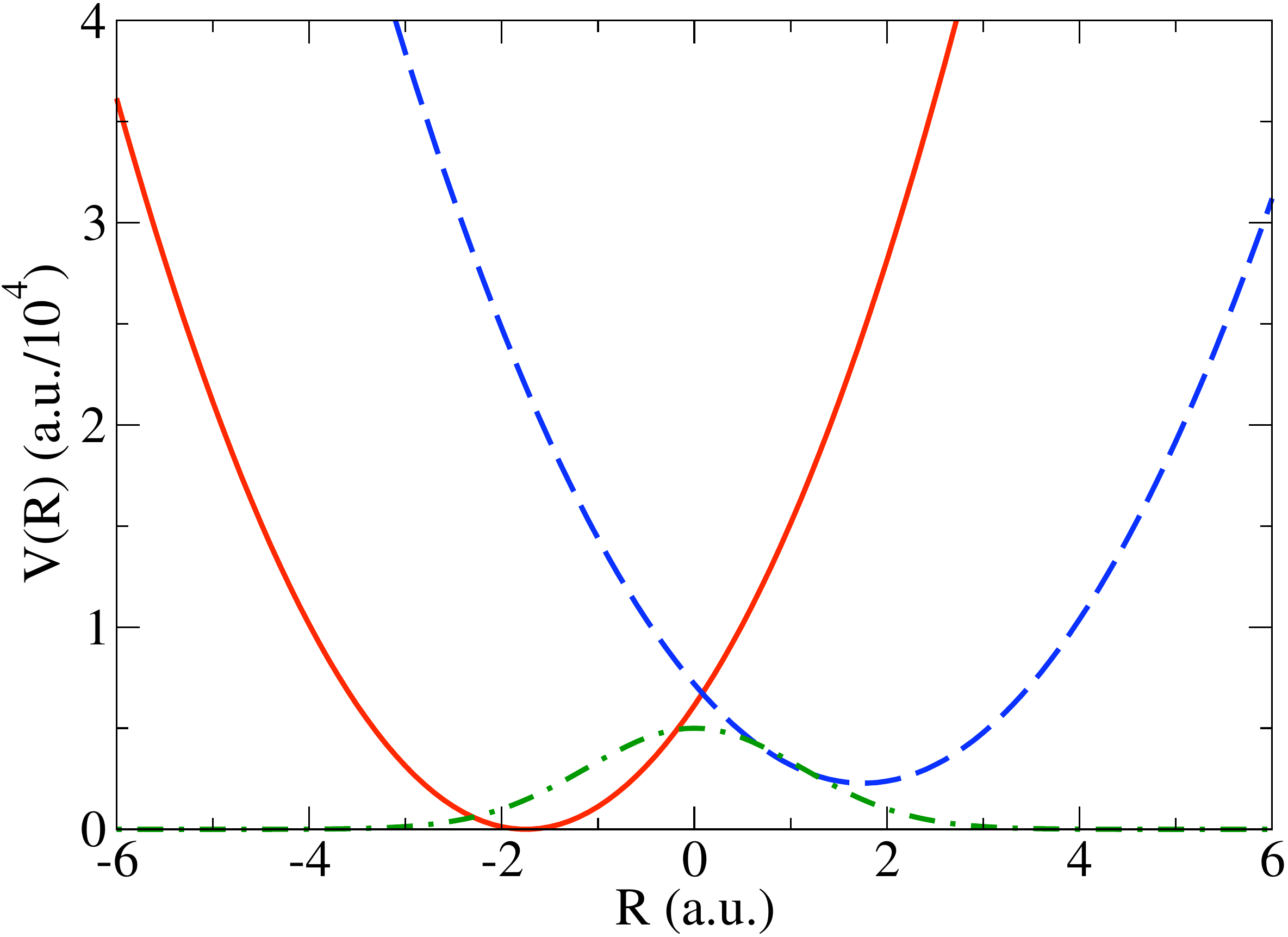}
	\caption{
	Diabatic potential energy curves for model I, 
	with state 1 in red (solid line), state 2 in blue (dashed line),
	and the coupling in green (dash-dotted line)
}
%  \caption{Diabatic potential energy curves for Model I as a function of nuclear position shown here in red and blue. 
%  The coupling is shown in green.}
  \label{fig:potential_twostate}
  \vspace{0cm}
\end{figure}
\end{center}

The matrix elements of the diabatic potential operator
are
\begin{eqnarray}
	\nonumber
	V_{ii}&=&\half k_i (R-R_i)^2+\epsilon_i\;\;\; \text{and}\\
	V_{ij}&=&c e^{-\alpha(R-R_{ij})^2},
	\label{eq:twostate_diab}
\end{eqnarray}
where the potential energy parameters are specified 
in Table \ref{table:twostate_potential}.
The simulation is performed with a nuclear
mass of $3600$ a.u. and temperature 
$T=8$ K. The potential energy curves for Model I
are plotted in Fig.~\ref{fig:potential_twostate}. 

\begin{table}[!htb]
	\caption{Parameters for Model I}
	\begin{tabular}{|c|c|}
		\hline
		Parameter & Value (in a.u.) \\
		\hline
		$k_1$ & 4 x $10^{-5}$ \\
		$k_2$ & 3.2 x $10^{-5}$ \\
		$R_1$ & -1.75 \\
		$R_2$ & 1.75 \\
		$\epsilon_1$ & 0 \\
		$\epsilon_2$ & 2.28 x $10^{-5}$ \\
		c & 5 x $10^{-5}$ \\
		$\alpha$ & 0.4 \\
		$R_{12}$ & 0 \\
		\hline
	\end{tabular}
	\label{table:twostate_potential}
\end{table}

\begin{center}
\begin{figure}[!htb]
	\includegraphics[angle=0,scale=0.4]{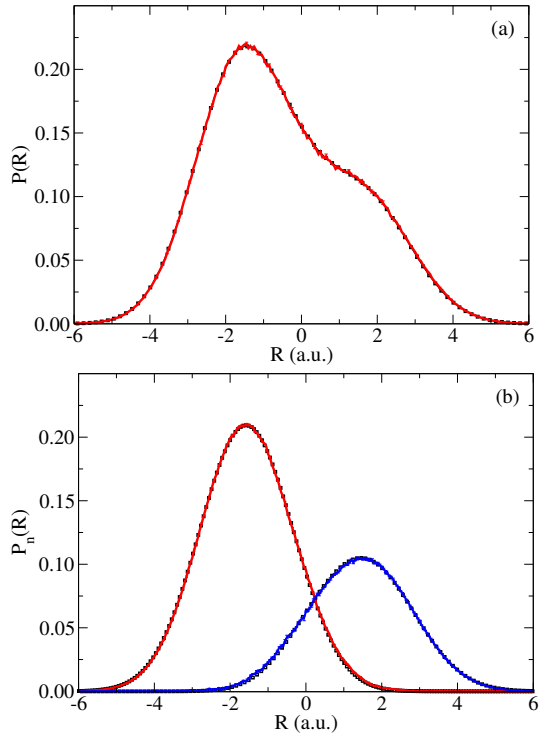}
	\caption{{\bf (a)} The nuclear probability distribution for model I,
	obtained using a 32-bead PI-ST simulation (red, solid line) and an 
	exact grid calculation (black squares). 
	{\bf (b)} The state-specific nuclear probability distribution 
	obtained from the PI-ST simulation, with state 1 in red (solid line,
	left peak) and state 2 in blue (solid line, right peak);
	black squares correspond to the exact results.
        }
	\label{fig:eqbm_twostate}
	 \vspace{0cm}
\end{figure}
\end{center}

The nuclear probability distribution [Eq.~(\ref{eq:nuc_prob_est})] 
obtained from a 32 bead PI-ST simulation is shown in 
Fig.~\ref{fig:eqbm_twostate}(a) and is graphically 
indistinguishable from a numerically exact
discrete variable representation (DVR)\cite{dtc92} grid calculation.
The tight convergence in this plot was achieved
using $5\times10^9$ Monte Carlo (MC) steps, and a similar number of steps
was found to be necessary for the corresponding PIMC calculation
in the discrete diabatic-state representation.
In Fig.~\ref{fig:eqbm_twostate}(b), we show that
the state-specific nuclear probability distribution
from this simulation also reproduces the exact results.
We further calculate the average total energy for Model I
and show, in Table \ref{table:avge_twostate}, that
the PI-ST result approaches the exact value in the limit
of a large number of beads.

% two state nuclear position distribution
%\vspace*{10cm}

% two state : energy
\begin{table}[!htb]
	\caption{Average Energy for Model I at $T=8$ K}
	\begin{tabular}{|c|c|} 
		\hline
		No: of Beads & Energy ($10^{-5}$ a.u.)\\
		\hline
		8 & 5.09 \\
		16 & 5.12 \\
		32 & 5.14 \\
		Exact & 5.145 \\
		\hline
	\end{tabular}
	\footnotetext{Statistical error for all cases
	is less than $10^{-7}$ a.u.}
	\label{table:avge_twostate}
\end{table}
%%fakesection model 2 results

Model II is a three-state system coupled to a single vibrational DoF.
It is based on a model used to simulate ultrafast photoinduced 
electron-transfer.\cite{de04}
The model includes a ground (G) electronic state, a locally excited (LE)
state that is accessible via photoexcitation, and 
a charge transfer (CT) state that facilitates
radiationless decay to the ground state. 
The CT state acts as a bridge state between the ground
and LE states, and it is coupled to both these states via a 
constant potential; there is no direct coupling
between the ground and LE states.

% three state : potential
%\vspace*{10cm}
\begin{center}
\begin{figure}[!htb]
	\includegraphics[angle=0,scale=0.35]{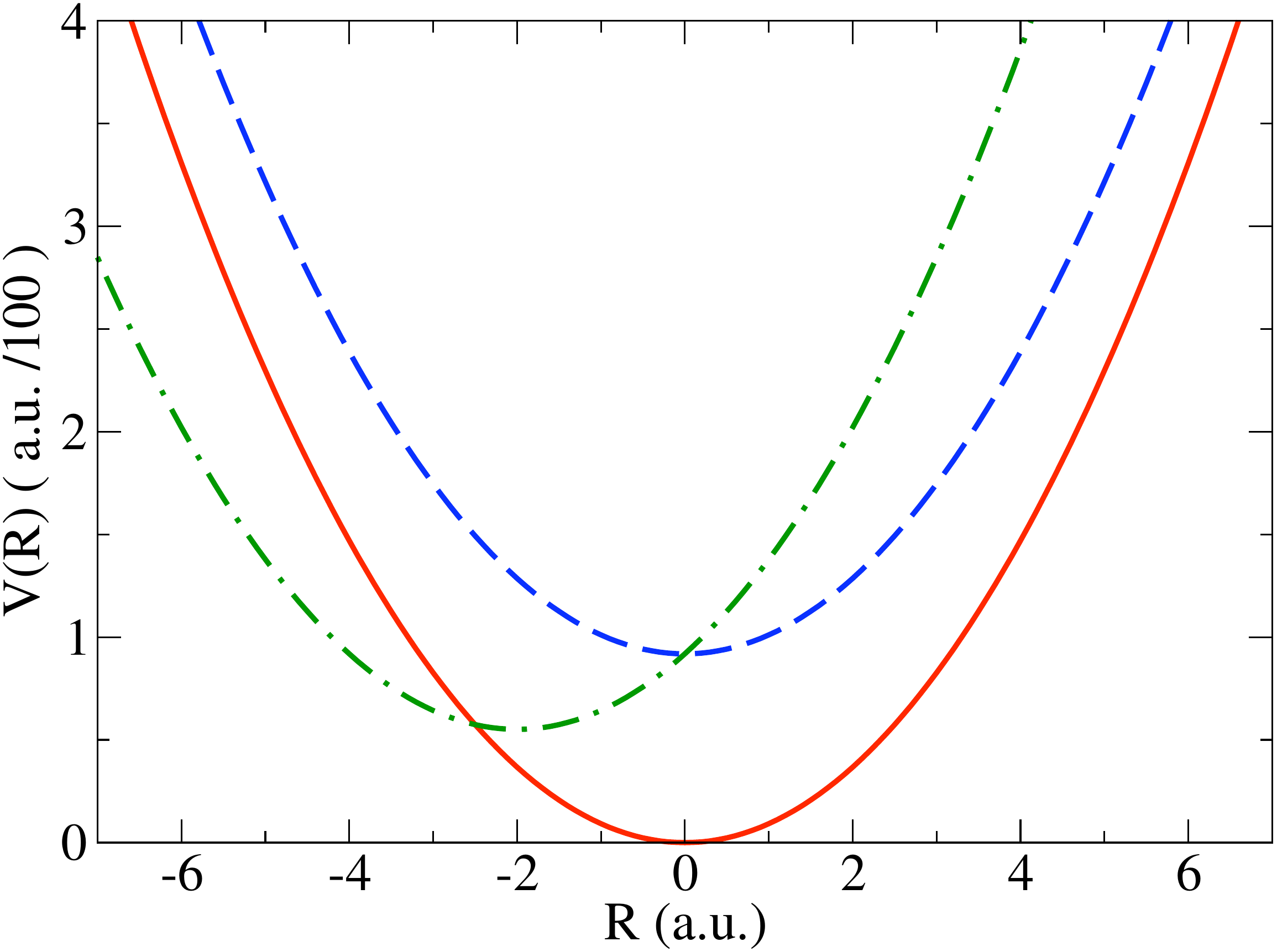}
	\caption{
	Diabatic potential energy curves for model II,
	with the G state in red (solid line), the LE state in 
	blue (dashed line) and the CT state in green (dash-dotted line);
	the constant coupling elements are not shown.}
   \label{fig:potential_threestate}
  \vspace{0cm}
\end{figure}
\end{center}

The matrix elements of the diabatic potential operator are
\begin{eqnarray}
	\nonumber
	V_i &=& \half \omega_s R^2 + k_i R + \epsilon_i\;\;\;\text{and} \\
	V_{ij} &=& c_{ij},
\end{eqnarray}
where $i,j\in\{\text{G},\text{CT},\text{LE}\}$, and the nuclear mass is $544.23$~a.u.
The potential energy parameters for Model II are provided in 
Table \ref{table:threestate_potential}, and the diabatic 
three-state potential is shown in Fig.~\ref{fig:potential_threestate}. 
The simulation is performed at $T=1500$ K, chosen 
such that all three states are thermally accessible.
\begin{table}[!t]
	\caption{Parameters for Model II}
	\begin{tabular}{|c|c|}
		\hline
		Parameter & Value (in eV) \\
		\hline
		$\omega_s$ & 0.05\\
		$k_G$ & 0 \\
		$k_{CT}$ & 0.1 \\
		$k_{LE}$ & 0 \\
		$\epsilon_G$ & 0 \\
		$\epsilon_{CT}$ & 0.25 \\
		$\epsilon_{LE}$ & 0.25 \\
		$c_{G,CT}$ & 0.02 \\
		$c_{CT,LE}$ & 0.03 \\
		$c_{G,LE}$ & 0 \\
		\hline
	\end{tabular}
	\label{table:threestate_potential}
\end{table}

\begin{center}
\begin{figure}[!htb]
	\includegraphics[angle=0,scale=0.4]{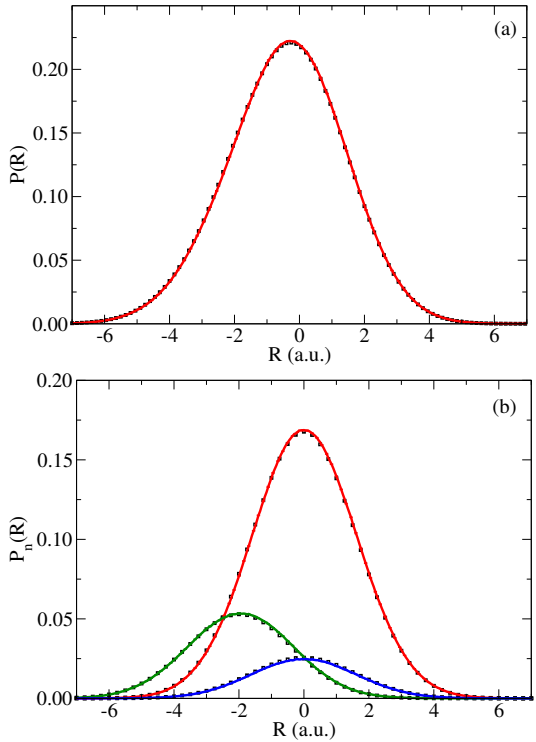}
	 \caption{{\bf (a)} The nuclear probability distribution
	 for model II, obtained using a 4-bead PI-ST simulation 
	 (red, solid line) and an exact grid calculation (black squares).
	 {\bf(b)} The state-specific nuclear probability distribution 
	 obtained from the PI-ST simulation with the G state, the CT 
	 state and the LE state in red, green and blue (solid lines,
	 in order of decreasing population) respectively. 
        }
	\label{fig:eqbm_threestate}
	\vspace{0cm}
\end{figure}
\end{center}

The converged nuclear probability distribution for this model is 
obtained from a 4-bead calculation and is graphically 
indistinguishable from the exact results from a DVR grid 
calculation, as seen in Fig.~\ref{fig:eqbm_threestate}(a).
These results were obtained using $10^9$ MC steps.
The state-specific nuclear probability distributions shown
in Fig.~\ref{fig:eqbm_threestate}(b) also reproduce the exact
results.

% three state nuclear position distribution
%\vspace*{10cm}

Further, in Table \ref{table:avge_threestate},
the results of the average energy calculation are reported, and
the exact results are recovered with increasing bead numbers.
% three state : energy
\begin{table}[!htb]
	\caption{Average Energy for Model II at T$=1500$ K}
	\begin{tabular}{|c|c|}
		\hline
		No: of Beads & Energy ($10^{-3}$ a.u.)\\
		\hline
		1 & 5.54 \\
		2 & 6.63 \\
		4 & 6.69 \\
		Exact & 6.688 \\
		\hline
	\end{tabular}
	\footnotetext{Statistical error in all cases is 
	less than $10^{-5}$ a.u.}
	\label{table:avge_threestate}
\end{table}

The equilibrium properties calculated for these
model systems demonstrate that the PI-ST
representation provides a general and exact 
statistical description of electronically 
nonadiabatic systems.
%%fakesection : TCFs
\section{Thermal Correlation Functions}
The PI-ST representation provides a natural means
to initialize SC trajectories to an exact quantum Boltzmann 
distribution for the calculation of real-time TCFs.
In this paper, we demonstrate this using the SC-IVR method,
which has already been successfully implemented in the mapping
framework.\cite{xs97,xs98,na07} 
% implemented in the mapping framework?
However, any trajectory-based model for real-time dynamics
could be combined with our exact PI-ST formulation.

A general real-time TCF is expressed as
\begin{equation}
	C_{AB}(t)=\frac{1}{Z}\text{Tr}\left[e^{-\beta\hop}A
	e^{i\hop t}Be^{-i\hop t}  \right],
	\label{eq:thermal_corr}
\end{equation}
where $A$ and $B$ are generic operators.
Substituting the PI-ST representation for the Boltzmann
operator from Eq.~(\ref{eq:pist}), the TCF can be written
\begin{eqnarray}
	\nonumber
	&&
	C_{AB}(t) = \frac{1}{Z}\int d\left\{\rvec_\alpha \right\}
	\int d\left\{\xvec_\alpha\right\}
	\prod_{\alpha=1}^{P-1}
	\mathcal{A}_\alpha \mathcal{G}_\alpha \mathcal{F}_\alpha
	\label{eq:pist_tcf}\\
	%\nonumber
	&&\times\;\bra{\xvec_P,\rvec_P}
	e^{-\bytwo{\beta_P\hop}}
	Ae^{i\hop t}Be^{-i\hop t}
	e^{-\bytwo{\beta_P\hop}}
	\proj\ket{\xvec_1,\rvec_1},\;\;\;
\end{eqnarray}
where $\mathcal{A}_\alpha$, $\mathcal{G}_\alpha$, and 
$\mathcal{F}_\alpha$ are defined in 
Eqs.~(\ref{eq:formula_a}-\ref{eq:formula_g}).
%%fakesection : HK-IVR
\subsection{Herman-Kluk (HK) IVR}
The HK-IVR propagator,\cite{mfh84,whm01,sad06,kgk06} 
\begin{equation}
	e^{-i\hop t} = \left(2\pi\right)^{-(N+f)}
	\int d\zvec_0\;\ket{\zvec_t}C^{\text{HK}}_t(\zvec_0)
	e^{iS_t(\zvec_0)}\bra{\zvec_0},
	\label{eq:hkivr}
\end{equation}
is a coherent state approximation to the full coordinate 
state SC-IVR propagator.
Here, $\ket{\zvec_0}=\ket{\xveco,\pveco}\ket{\rveco,\Pveco}$
represents the initial electronic and nuclear coherent states
of width $\gamma$ and $\Gamma$, respectively,
and $\ket{\zvec_t}$ is obtained from classically time-evolving the 
initial positions and momenta for time $t$.
In addition, $S_t$ is the classical action, and the HK prefactor 
is given by
\begin{equation}
   C_t^{\text{HK}}(\zvec_0) = \text{Det}\left[
   \half\;\mathbf{g}^T\;\frac{\partial\zvec_t}{\partial\zvec_0}
   \;\mathbf{g^{-1}}\;
   \right]^\half ,
   \label{eq:hkpref}
\end{equation}
where $\mathbf{g}=\left( (\gamma,\Gamma)^{1/2}\;,\,
i(\gamma,\Gamma)^{-1/2} \right)$.

The forward and backward propagators in Eq.~(\ref{eq:pist_tcf})
can be replaced by HK-IVR propagators to obtain an expression 
with a double phase-space integral over initial conditions,
\begin{eqnarray}
	\nonumber
	\lefteqn{
	C_{AB}^\text{HK}(t)=
	\frac{\left( 2\pi \right)^{-2(N+f)}}{Z}\int d\left\{\rvec_\alpha \right\}
	\int d\left\{\xvec_\alpha\right\}
	\prod_{\alpha=1}^{P-1}
	\mathcal{A}_\alpha \mathcal{G}_\alpha \mathcal{F}_\alpha
	}
	\\
	\nonumber
	&\times&
	\int d\zvec_0\int d\zvec_0^\prime\;
	e^{i[ S_{-t}(\zvec_0^\prime) + 
	S_t(\zvec_0) ]}
	C_{-t}(\zvec_0^\prime)
	C_t(\zvec_0)
	\bra{\zvec_t^\prime}B\ket{\zvec_t}
	%&\times&\prod_{\alpha=1}^{P-1}\bra{\xvec_\alpha,\rvec_\alpha}
	%e^{-\beta_P\hop}\proj
	%\ket{\xvec_{\alpha+1}\rvec_{\alpha+1}}
	\\
	&\times&
	\bra{\xvec_P,\rvec_P}e^{-\beta_P\hop/2}A\ket{\zvec_0^\prime}
	\bra{\zvec_0}
	e^{-\beta_P\hop/2}\proj\ket{\xvec_1,\rvec_1}.
	\label{eq:hkivr_tcf}
\end{eqnarray}
MC integration of the resulting oscillatory
integrand is known to be challenging,\cite{whm01}
and despite several advances in the evaluation of such
integrands,\cite{nm87,xs99,re00,vj09} the HK-IVR 
approach is limited to systems with few DoF.
Nonetheless, we include the HK-IVR implementation 
to illustrate the generality of our exact PI
initialization approach and to provide a reference 
semiclassical result to compare against the linearized IVR 
implementation.
%%fakesection : LSC-IVR
\subsection{Linearized IVR (LSC-IVR)}
The LSC-IVR approximation to the coordinate state SC-IVR 
expression for correlation functions is obtained from
a first order expansion of the difference in 
the actions of the forward and backward trajectories.
\cite{xs97l,xs98}
The resulting expression corresponds to the classical
Wigner model and can be written
\begin{eqnarray}
	C_{AB}^{\text{LSC}}&(t)&=\\
	\nonumber
	&&
	\frac{\left(2\pi\right)^{-(N+f)}}{Z}
	\int d\pvec_0 \int d\xvec_0 
	A_\text{W}^\beta(\xvec_0,\pvec_0)
	B_\text{W}(\xvec_t,\pvec_t),
	\label{eq:thermal_corr_lsc}
\end{eqnarray}
where $A_\text{W}^\beta=\left( e^{-\beta\hop}A \right)_\text{W}$, and 
the Wigner transformed operators are obtained by evaluating
\begin{equation}
	O_\text{W}(\xvec,\pvec)=\int d\Delta\xvec \bra{\xvec-\frac{\Delta\xvec}{2}}
	O\ket{\xvec+\frac{\Delta\xvec}{2}}e^{i\pvec^T\cdot\Delta\xvec}.
\end{equation}
The LSC-IVR approximation largely
fails to capture quantum coherence effects,\cite{xs98,whm01,na07} 
but it successfully describes other quantum effects 
such as zero point energy and tunneling, making it
suitable for many condensed phase applications.
\cite{whm01,jl06,jl08,jl09}

The PI-ST representation of the Boltzmann operator can be
substituted in the expression for the TCF  in 
Eq.~(\ref{eq:thermal_corr_lsc}) to obtain
\begin{eqnarray}
	\nonumber
	C_{AB}^\text{LSC}(t)&=&\frac{\left( 2\pi \right)^{-(N+f)}}{Z}
	\int d\left\{\rvec_\alpha\right\}
	\int d\left\{\xvec_\alpha\right\}
	\prod_{\alpha=1}^{P-1}\mathcal{A}_\alpha 
	\mathcal{G}_\alpha \mathcal{F}_\alpha
	\\
	&&\hspace{0.8in}\times\;
	\int d\zvec_0 \tilde{A}^\beta_\text{W}(\zvec_0) 
	B_\text{W}(\zvec_t),
	\label{eq:lsc_tcf}
\end{eqnarray}
where
\begin{eqnarray}
	\nonumber
	\lefteqn{\tilde{A}^\beta_W(\zvec_0)=
	\int d\Delta \xvec \int d\Delta\rvec\;
	e^{i\pveco\cdot\Delta\xvec + i\Pveco\cdot\Delta\rvec}
	}
	\\
	\nonumber
	&\times&
	\bra{\xveco-\bytwo{\Delta\xvec},\rveco-\bytwo{\Delta\rvec}}
	e^{-\bytwo{\beta_P}\hop}\proj\ket{\xvec_1,\rvec_1}\\
        &\times&
	\bra{\xvec_P,\rvec_P}
	e^{-\bytwo{\beta_P}\hop}A
	\ket{\xveco+\bytwo{\Delta\xvec},\rveco+\bytwo{\Delta\rvec}}.
	\label{eq:atilde_wigner}
\end{eqnarray}

We recognize that using the exact PI-ST representation of
the Boltzmann operator introduces an oscillatory term in 
the LSC-IVR formulation via $A^\beta_\text{W}(\zvec_0)$.
For future applications to large systems, this oscillatory term
can be eliminated using techniques such as the 
Thermal Gaussian Approximation,\cite{paf04} 
for the Boltzmann matrix elements in Eq.~(\ref{eq:atilde_wigner}).
Since the remaining $(P-1)$ Boltzmann terms in Eq.~(\ref{eq:lsc_tcf})
will still be treated using the exact
PI-ST representation, we expect that this approach would 
introduce only small deviations from the exact quantum 
statistical description of a nonadiabatic system.
%%fakesection : Implementation
\subsection{Implementation of PI-ST initialization}
Equations (\ref{eq:hkivr_tcf}) and (\ref{eq:lsc_tcf}) can
both be expressed in the form
\begin{eqnarray}
	\nonumber
	\lefteqn{
	C_{AB}^\xi(t)=\frac{1}{Z}
	\int d\left\{\rvec_\alpha\right\}
	\int d\left\{\xvec_\alpha\right\}
	W(\{\xvec_\alpha\},\{\rvec_\alpha\})
	}\\
	\nonumber
	&&\hspace{0.7in}
	\times\;\;f(\{\xvec_\alpha\},\{\rvec_\alpha\})
	\Phi^{\xi}(\xvec_1,\xvec_P,\rvec_1,\rvec_P,t)
	\\
	\nonumber
	&& \vspace{1in}=
	\left<\;\Phi^{\xi}(\xvec_1,\xvec_P,\rvec_1,\rvec_P,t) f(\{\xvec_\alpha\},\{\rvec_\alpha\})\;
	\right>_W
	\\
	&&\hspace{1.45in}/
	\left<\;f_Z(\{\xvec_\alpha\},\{\rvec_\alpha\})\;
	\right>_W,
	\label{eq:alg_tcf}
\end{eqnarray}
where $W(\{\xvec_\alpha\},\{\rvec_\alpha\})$ is a
sampling distribution and 
$f(\{\xvec_\alpha\},\{\rvec_\alpha\})$ and 
$f_Z(\{\xvec_\alpha\},\{\rvec_\alpha\})$ are
weighting factors, all of which emerge from the PI-ST treatment
of the Boltzmann operator.
The term $\Phi^\xi(\xvec_1,\rvec_1, \xvec_P, \rvec_P, t)$ in 
Eq.~(\ref{eq:alg_tcf}) contains the real-time information 
obtained from the SC trajectories, and the superscript 
$\xi\in\{\text{HK,LSC}\}$ indicates which SC approximation 
is employed.

The calculation of the TCF in Eq.~(\ref{eq:alg_tcf}) is
performed by first generating an ensemble of configurations 
from the probability distribution $W(\{\xvec_\alpha\},\{\rvec_\alpha\})$. 
Then, as in standard SC-IVR calculations,\cite{whm01} MC
importance sampling is used to evaluate $\Phi^\xi(\xvec_1,\rvec_1,
\xvec_P,\rvec_P,t)$.
For the HK-IVR implementation,
\begin{eqnarray}
	\lefteqn{\Phi^{\text{HK}}(\xvec_1,\xvec_P,\rvec_1,\rvec_P,t)
	= }\\
	\nonumber
	&&\;\;N^\text{HK}(t)\left<
	\phi^\text{HK}(\zvec_0,\zvec_t,\zvec_0^\prime,
	\zvec_t^\prime;\xvec_1,\xvec_P,\rvec_1,\rvec_P)
	\right>_{
	\Pi^\text{HK}, %(\zvec_0,\zvec_0^\prime;\rvec_1,\rvec_P),
	}
	\label{eq:hk_ivr}
\end{eqnarray}
and for the LSC-IVR implementation,
\begin{eqnarray}
	\lefteqn{\Phi^{\text{LSC}}(\xvec_1,\xvec_P,\rvec_1,\rvec_P,t)
	= }\\
	\nonumber
	&&\;\;\;\;\;\;N^\text{LSC}(t)\left<
	\phi^\text{LSC}(\zvec_0,\zvec_t
	;\xvec_1,\xvec_P,\rvec_1,\rvec_P)
	\right>_{
	\Pi^\text{LSC}%(\zvec_0;\rvec_1,\rvec_P)
	},
	\label{eq:lsc_ivr}
\end{eqnarray}
where $\Pi^\xi$ is a probability 
distribution function used to generate an ensemble of 
initial coordinates and momenta for the SC trajectories,
$\phi^\xi$ is the corresponding time-dependent estimator,
and $N^\xi(t)$ is the associated time-dependent normalization term. %constant.

In this paper we calculate the electronic state population TCF, 
$C_{nn}^\xi(t)$, where $A=B=\ket{n}\bra{n}$ in 
Eq.~(\ref{eq:thermal_corr}). For this special case, the detailed
form of the functions described above are provided in the appendix.
%%fakesection : {Results Model III}
\subsection{Dynamics simulation results}
We calculate the real-time electronic state population TCF
using the PI-ST representation of the Boltzmann
distribution to initialize SC trajectories for
dynamics in the LSC-IVR and HK-IVR frameworks.
The first set of results presented are for
model III, a simple two-state system
described by the Hamiltonian,
\begin{equation}
	H = \alpha\sigma_z + \Delta\sigma_x,
\end{equation}
where $\sigma_z$ and $\sigma_x$ are the Pauli matrices.
The potential parameters are 
$\left(\alpha,\Delta\right)~=~\left( 0.5,1 \right)$ in a.u.
The mapping Hamiltonian for a two-state system assumes
a quadratic form for which both the HK-IVR and LSC-IVR
formulations are exact.
The simulation is performed at a reciprocal temperature
of $\beta~=~1$~a.u., using coherent states of width 
$\gamma~=~1$~a.u. SC trajectories are integrated
using an Adams-Bashforth-Moulton predictor-corrector integrator.\cite{gh76}
Exact results are obtained from a DVR grid calculation.

\begin{center}
\begin{figure}[!htb]
	\includegraphics[angle=0,scale=0.335]{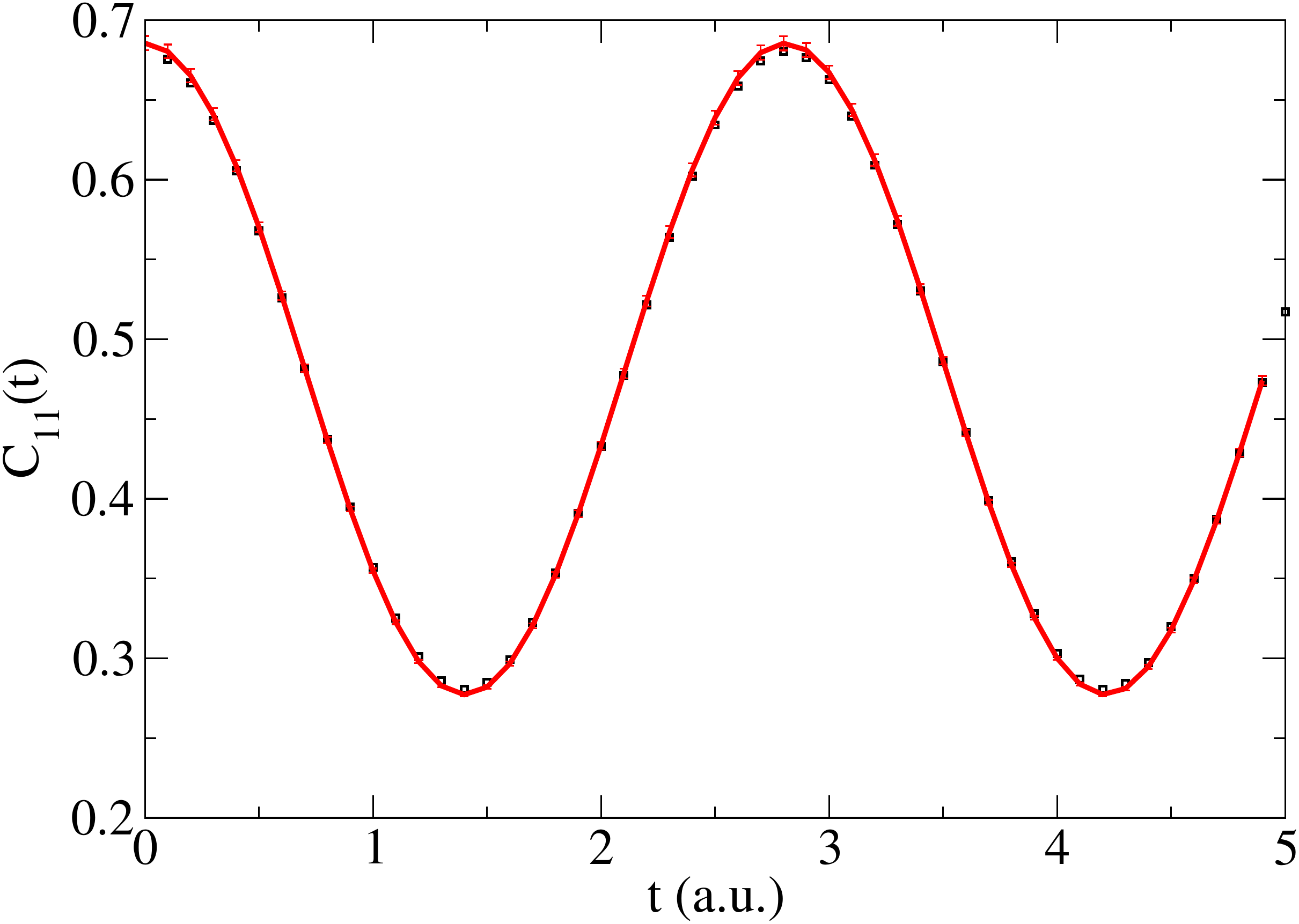}
	%\subfigure{
	%\includegraphics[angle=0,scale=0.4]{./model1_cpp_lsc_hk.pdf}
        %\label{fig:model1_tcorr}
	%}
	%\subfigure{
	%}
	%\subfigure{
	%\includegraphics[angle=0,scale=0.4]{./model2dt_cpp.pdf}
        %\label{fig:model2dt_tcorr}
	%}
	 \caption{The real-time electronic state population TCF for model  
III, obtained
   from the LSC-IVR method with PI-ST initialization (red, solid line) and an  
exact grid calculation (black squares).
   Graphically indistinguishable results were also obtained using the  
HK-IVR method with PI-ST
   initialization.
   }
   \label{fig:model2_tcorr}
  \vspace{0cm}
\end{figure}
\end{center}

Fig.~\ref{fig:model2_tcorr} illustrates that the $C_{11}(t)$ TCF 
calculated using the LSC-IVR implementation
reproduces the exact results, as expected. 
Simulations performed using the HK-IVR implementation yielded 
graphically indistinguishable results.
These calculations were performed using an 8-bead simulation with
$10^8$ MC steps for the sampling of the probability distribution
$W(\{\xvec,\rvec\})$. For each of these configurations,
ensembles of 120 trajectories and 10 trajectories were generated
to obtain the HK-IVR and LSC-IVR TCFs, respectively.
% Two state : thermal correlation function (population-population) Model4
%\vspace*{10cm}

%%fakesection : Results Model IV
Model IV is a two-state system coupled to a single 
nuclear DoF of mass $M~=~1$~a.u. 
The Hamiltonian for this model is
\begin{equation}
	H = \frac{P^2}{2M} + \half R^2 + \alpha R\sigma_z + \Delta\sigma_x,
\end{equation}
with parameters (in a.u.) $\alpha~=~1$, $\Delta~=~1$,
and $\beta~=~1$.
We use coherent states of width $\gamma~=~1$~a.u and 
$\Gamma~=~1$~a.u. for the electronic and nuclear 
DoF, respectively and 4-bead simulations were performed with $10^8$ MC
steps for sampling the probability distribution 
$W(\{\xvec,\rvec\})$.

\begin{center}
\begin{figure}[!htb]
	\includegraphics[angle=0,scale=0.35]{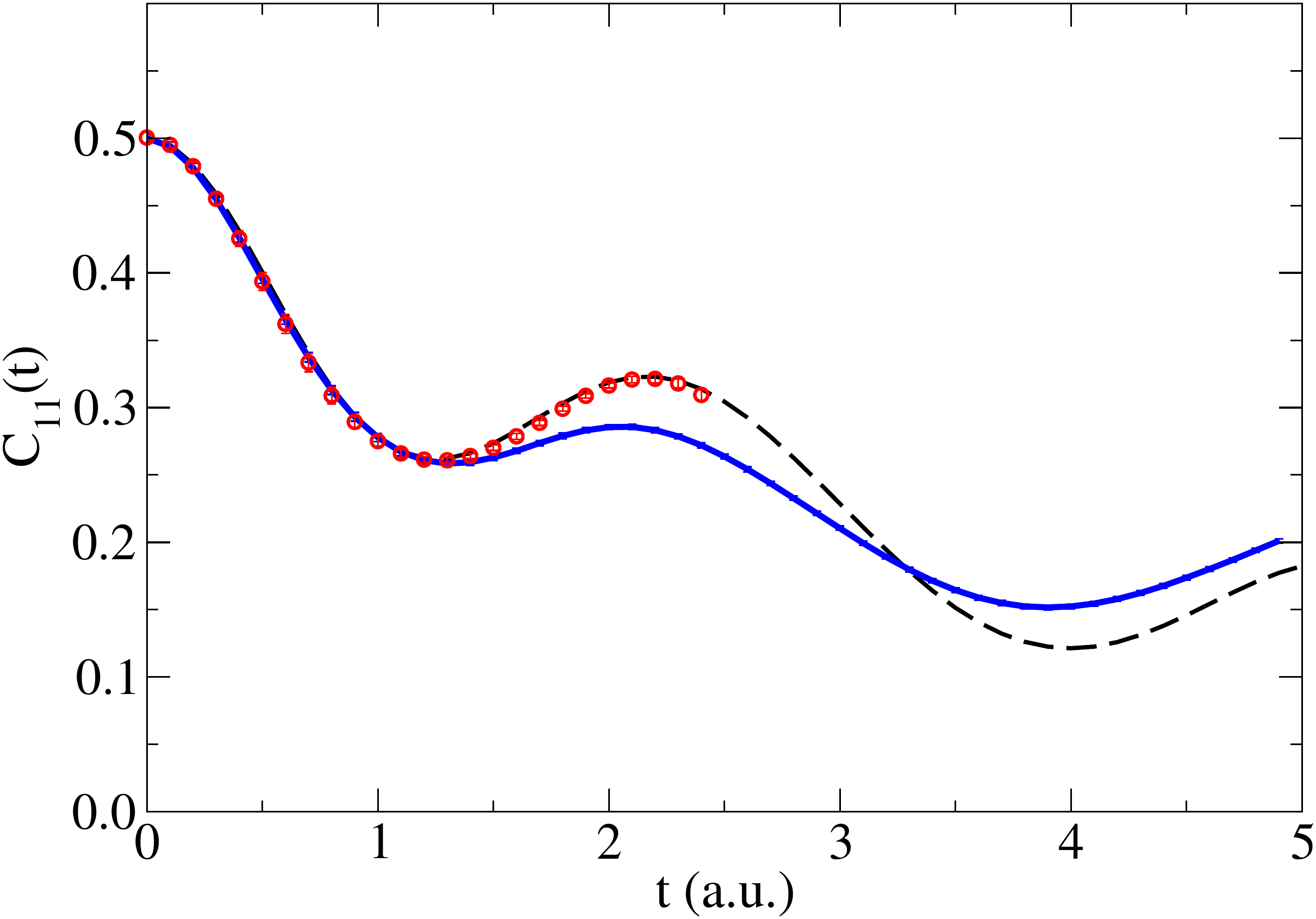}
	\caption{The real-time electronic state population TCF for model  
IV, obtained
   from the HK-IVR method with PI-ST initialization (red, dotted line), the  
LSC-IVR method with PI-ST initialization (blue, solid line), and an exact  
grid calculation (black, dashed line).}
	   \label{fig:withnuc_corr}
  \vspace{0cm}
\end{figure}
\end{center}

In Fig.~\ref{fig:withnuc_corr}, the TCFs from the HK-IVR and 
LSC-IVR simulations are compared with the results from an exact
DVR grid calculation.
The HK-IVR implementation reproduces the exact results with 
remarkable accuracy, although the calculation required an
ensemble of $80,000$ trajectories per equilibrium configuration 
to converge results up to a time of $2.5$ a.u.; results for 
longer times would require even larger numbers of trajectories.
The LSC-IVR simulation, however, required only $8000$ trajectories
per equilibrium configuration to achieve convergence up to $5$ a.u. in time.
As expected, the LSC-IVR approximations dampens the oscillations 
seen in the exact calculation.
% Two state one mode :: thermal correlation function (population-population) 
\vspace*{-0.1cm}
\vspace{-0.75cm}

%%fakesection : conclusions
\section{Conclusions}
We have derived an exact PI-ST representation for the Boltzmann
statistics of N-level systems using continuous path variables for 
both the electronic and nuclear DoF.
This result is demonstrated 
to be numerically exact for equilibrium 
simulations of two- and three-state systems. 
Additionally, the PI-ST representation is used to initialize
trajectories in the SC-IVR framework allowing for the calculation
of real-time TCFs with encouraging accuracy.
Natural future applications of this methodology include
charge transfer reactions in the condensed phase and metal-surface
energy transfer processes, for which excited electronic states are
thermally accessible.
%The PIMMST expression also provides a potential
%step towards constructing model dynamics that 
%preserve the quantum Boltzmann distribution in 
%time and provide accurate long-time information.
\section{Acknowledgements}
The authors sincerely thank Bill Miller for valuable comments and
insights.  This work was partially supported by a STIR grant from the Army
Research Office.  T.F.M. additionally acknowledges support from a Camille
and Henry Dreyfus Foundation New Faculty Award and an Alfred P. Sloan
Foundation Research Fellowship.
\begin{appendix}
\section{Electronic state population TCF}
The detailed form for all functions used to calculate 
$C_{nn}^\xi(t)$ are provided here.
The functions $W(\{\xvec_\alpha\},\{\rvec_\alpha\})$,
$f(\{\xvec_\alpha\},\{\rvec_\alpha\})$, and 
$f_Z(\{\xvec_\alpha\},\{\rvec_\alpha\})$ arise
from the PI-ST representation of the Boltzmann operator and are 
identical for both the LSC-IVR and HK-IVR implementations.
Specifically,
\begin{eqnarray}
   \nonumber
   W(\{\xvec_\alpha\},\{\rvec_\alpha\})&=&e^{-\beta_PV_0(\rvec_P)/2}
   \;\mathcal{G}_P\;|\mathcal{F}_P|\\
   &\times& \prod_{\alpha=1}^{P-1} e^{-\beta_PV_0(\rvec_\alpha)}
   \;\mathcal{A}_\alpha\;\mathcal{G}_\alpha\;|\mathcal{F}_\alpha|,\;\;\;
   \label{eq:w_ivr}
\end{eqnarray}
where $\mathcal{A}_\alpha$, $\mathcal{G}_\alpha$, and 
$\mathcal{F}_\alpha$ are defined in 
Eqs.~(\ref{eq:formula_a}-\ref{eq:formula_g});
\begin{equation}
	f(\{\xvec_\alpha\},\{\rvec_\alpha\})= 
	\frac{e^{-\beta_PV_0(\rvec_1)/2}\text{sgn}(\mathcal{F})}
	{\xvec_P^T\intmat(\rvec_P)\xvec_1},
	\label{eq:f_ivr}
\end{equation}
where $\text{sgn}(\mathcal{F})$ is defined in Eq.(\ref{eq:signf}) and
the elements of $\intmat$ defined in Eq.~(\ref{eq:int_mat});
and 
\begin{equation}
   f_Z\left(\{\mathbf{x}_\alpha\},\{\mathbf{R}_\alpha \} \right) = 
   \text{sgn}(\mathcal{F})e^{-\frac{MP}{2\beta}(\mathbf{R}_P-\mathbf{R}_1)^T
   \cdot (\mathbf{R}_P-\mathbf{R}_1)} \; e^{-\beta_P V_0(\mathbf{R}_P)/2 }
\label{eq:fz_ivr}
\end{equation}

The terms required to evaluate $\Phi^\xi(\xvec_1,\rvec_1,\xvec_P,\rvec_P, t)$ 
in Eqs.~(\ref{eq:hk_ivr}) and 
(\ref{eq:lsc_ivr}) are derived by 
substituting $A=B=\ket{n}\bra{n}$ into 
Eqs.~(\ref{eq:hkivr_tcf}) and (\ref{eq:lsc_tcf}).
In the HK-IVR framework, this yields, the probability distribution function
\begin{eqnarray}
	\nonumber
	\lefteqn{
	\Pi^{\text{HK}}(\zvec_0,\zvec_0^\prime;\rvec_1,\rvec_P)
	=e^{-\frac{\beta}{\beta\Gamma+2MP}( \Pveco^T\cdot\Pveco + 
	\Pvecopt\cdot\Pvecop )}
	}
	\\
	\nonumber
	&&e^{
	-\frac{MP\Gamma}{\beta\Gamma+2MP}\left( (\rvecop-\rvec_P)^T\cdot
	(\rvecop-\rvec_P)+ (\rveco-\rvec_1)^T\cdot(\rveco-\rvec_1) \right)
	}
	\\
	&&
	e^{-\frac{\gamma}{2(\gamma+1)}( \xveco^T\cdot\xveco + \xvecopt\cdot\xvecop )
	-\frac{1}{2(\gamma+1)}( \pveco^T\cdot\pveco + \pvecopt\cdot\pvecop )},
	\label{eq:pi_hk}
\end{eqnarray}
and the corresponding estimator
\begin{eqnarray}
	\nonumber
	&\phi^{\text{HK}}&(\zvec_0,\zvec_0^\prime,\zvec_t,\zvec_t^\prime;
	\xvec_1,\xvec_P,\rvec_1,\rvec_P)=\nonumber\\
	\nonumber
	&&\left[\gamma_{0}\xvecp_t-i\pvecp_t\right]_n
	\left[\gamma_{0}\xvec_t+i\pvec_t\right]_n
	\\
	\nonumber
	&\times&
	\left[
	\xvec_P^T\intmat^\prime(\rvec_P)
	(\gamma\xvecop+i\pvecop)
	\right]_n
	\left(
	(\gamma_{0}\xveco-i\pveco)^T\intmat^\prime(\rvec_1)
	\xvec_1\right)
	\\
	\nonumber
	&\times&
	C_{-t}(\zvec_0^\prime)
	C_t(\zvec_0)
	e^{i[ S_{-t}(\zvec_0^\prime) + 
	S_t(\zvec_0) ]}
	\\
	\nonumber
	&\times&
	e^{\frac{i}{\gamma+1}\pveco^T\cdot\xveco
	-\frac{i}{\gamma+1}\pvecopt\cdot\xvecop
	-\frac{\Gamma}{4}(\rvecp_{t}-\rvec_t)^T.(\rvecp_t-\rvec_t)}
	\\
	\nonumber
	&\times&
	e^{
	-\frac{1}{4\Gamma}(\Pvecp_{t}-\Pvec_t)^T.(\Pvecp_t-\Pvec_t)
	+\frac{i}{2}(\Pvec_t+\Pvecp_{t})^T\cdot(\rvecp_{t}-\rvec_t)}
	\\
	\nonumber
	&\times&
	e^{\frac{2iMP}{\beta\Gamma+2MP}
	\left(\Pvecopt\cdot(\rvec_P-\rvecp_0)
	-\Pveco^T\cdot(\rvec_1-\rvec_0)\right)
	+\frac{i}{\gamma+1}\left(\pvecpt_{t}\cdot\xvecp_{t}
	-\pvec_t^T\cdot\xvec_t \right)}
	\\
	&\times&
	e^{-\frac{\gamma}{2(\gamma+1)}(\xvec_t^T\cdot\xvec_t+
	\xvecpt_{t}\cdot\xvecp_t)
	-\frac{1}{2(\gamma+1)}(\pvec_t^T\cdot\pvec_t+
	\pvecpt_t\cdot\pvecp_t)},
	\label{eq:phi_hk}
\end{eqnarray}
where the elements of the matrix $\intmat^\prime(\rvec)$ 
are identical to those of the matrix in Eq.~(\ref{eq:int_mat}) 
with $\beta~\rightarrow~\beta/2$.

In the LSC-IVR framework, the probability distribution is
\begin{eqnarray}
	\nonumber
	\Pi^{\text{LSC}}(\zvec_0;\rvec_1,\rvec_P)=\nonumber\\
	&&\hspace{-1in}e^{-\frac{2MP}{\beta}
	\left(\rveco-\half(\rvec_P+\rvec_1)\right)^T
	\cdot\left(\rveco-\half(\rvec_P+\rvec_1)\right)}\nonumber
	\\ 
%	&&\hspace{0.2in}\times\;
	&&\hspace{-1in}\times e^{-\frac{\beta}{2MP}\Pveco^T.\Pveco
	-\xveco^T\cdot\xveco
	-\pveco^T\cdot\pveco},
	\label{eq:pi_lsc}
\end{eqnarray}
and the corresponding estimator is
\begin{eqnarray}
	\nonumber
	\lefteqn{
	\phi^\text{LSC}(\zvec_0,\zvec_t;\xvec_1,\xvec_P,
	\rvec_1,\rvec_P)=
	\left(\left[ \xvec_t \right]_n^2+
	\left[ \pvec_t \right]_n^2-\half\right)\times
	}
	\\
	\nonumber
	&&\left(\left[
	\xvec_0+i\pvec_0\right]_n
	(\xvec_0-i\pvec_0)^T\intmat^\prime(\rvec_1)
	\xvec_1
	-\half\left[\intmat^\prime(\rvec_1)\xvec_1\right]_n
	\right)\\
	&&\hspace{0.4in}\times
	\left[\xvec_P^T\intmat^\prime(\rvec_P)\right]_n 
	e^{i\Pveco^T\cdot(\rvec_P-\rvec_1)
	-\xvec_t^T\cdot\xvec_t-\pvec_t^T\cdot\pvec_t}.
	\label{eq:phi_lsc}
\end{eqnarray}

A final step in the TCF calculation is to obtain $N^\xi(t)$, % the 
%normalization constant $N^\xi$,
 which is required to
both normalize the probability distribution
function $\Pi^\xi$ and to correct for the non-unitarity
of the SC propagator.\cite{eac01}
Enforcing that 
the total electronic state population is conserved 
at all times yields
\begin{eqnarray}
	N^\xi(t) = \frac{C^{\xi}_{nn}(0)}
	{\sum_{m=1}^N \tilde{C}^{\xi}_{nm}(t)},
	\label{eq:norm_ivr}
\end{eqnarray}
where $C_{nn}^\xi(0)=\text{Tr}\left[ e^{-\beta\hop}\proj_n \right]/Z$
can be calculated from an exact PI-ST equilibrium simulation.
The terms $\tilde{C}_{nm}^\xi(t)$ in Eq.~(\ref{eq:norm_ivr}) are
un-normalized TCFs, such that
${C}_{nm}^\xi(t)\equiv N^\xi(t)\tilde{C}_{nm}^\xi(t)$,
where $A=\ket{n}\bra{n}$ and $B=\ket{m}\bra{m}$ in 
Eq.~(\ref{eq:thermal_corr}).
These un-normalized terms are obtained following the steps
described above, except that the first line on the right-hand side
of Eqs.~(\ref{eq:phi_hk}) and (\ref{eq:phi_lsc})
is modified to include the 
$m^\text{th}$ (rather than the $n^\text{th}$) component of vectors $\xvec_t,\pvec_t,
\xvec_t^\prime,\;\text{and}\;\pvec_t^\prime$. 
%These un-normalized terms are obtained following the steps
%described above, except that the expressions for 
%$\phi^\xi$ are modified to include the 
%$m^\text{th}$ (rather than the $n^\text{th}$) component of vectors $\xvec_t,\pvec_t,
%\xvec_t^\prime,\;\text{and}\;\pvec_t^\prime$, 
%instead of the $n^\text{th}$ component in the first line
%of Eqs.~(\ref{eq:phi_hk}) and (\ref{eq:phi_lsc}).
The additional computational cost associated with this normalization term is negligible.
%calculation of a full set $\tilde{C}^\xi_{nm}(t)$ required for 
%the normalization involves negligible computational effort over 
%the calculation of $\tilde{C}_{nn}^\xi(t)$.
\end{appendix}
%%fakesection : bib
%% mixed quantum classical methods

\end{document}